REVIEW MANUSCRIPT

Synthetic and Systems Biotechnology Journal

# A review of DNA restriction-free overlapping sequence cloning techniques for synthetic biology

**Isabella Frighetto Bomfiglio**[1,2,*,†], **Isabelli Seiler de Medeiros Mendes**[1,2,*,†], and **Diego Bonatto**[1,2,*]

[1]Laboratório de Biologia Molecular e Computacional, Centro de Biotecnologia da UFRGS, Departamento de Biologia Molecular e Biotecnologia, Universidade Federal do Rio Grande do Sul, Porto Alegre, RS, Brazil.

[2]Bioprocess and Biotechnology for Food Research Center (Biofood), Food Science and Technology Institute (ICTA), Universidade Federal do Rio Grande do Sul, Av. Bento Gonçalves 9500, Porto Alegre, RS, 91501-970, Brazil.

**Short title:** DNA recombineering techniques.

**\*Corresponding author:**

Diego Bonatto

Centro de Biotecnologia da UFRGS - Laboratório 212

Departamento de Biologia Molecular e Biotecnologia

Universidade Federal do Rio Grande do Sul – UFRGS

Avenida Bento Gonçalves 9500 - Prédio 43421

Caixa Postal 15005

Porto Alegre – Rio Grande do Sul

BRAZIL

91509-900

Phone: (+55 51) 3308-9410

E-mail: diego.bonatto@ufrgs.br

Contract/grant sponsor: CNPq, FAPERGS

[†] Both authors contributed equally to the work.

**Abstract**

DNA cloning methods are fundamental tools in molecular biology, synthetic biology, and genetic engineering that enable precise DNA manipulation for various scientific and biotechnological applications. This review systematically summarizes the major restriction-free overlapping sequence cloning (RFOSC) techniques currently used in synthetic biology and examines their development, efficiency, practicality, and specific applications. *In vitro* methods, including Gibson Assembly, Circular Polymerase Extension Cloning (CPEC), Polymerase Incomplete Primer Extension (PIPE), Overlap Extension Cloning (OEC), Flap Endonuclease Cloning (FEN-Cloning), and commercially available techniques such as TOPO® and In-Fusion™, have been discussed alongside hybrid approaches such as Ligation-Independent Cloning (LIC), Sequence-Independent Cloning (SLIC), and T5 Exonuclease-Dependent Assembly (TEDA). Additionally, *in vivo* methods leveraging host recombination machinery, including Yeast Homologous Recombination (YHR), *In Vivo* Assembly (IVA), Transformation-Associated Recombination (TAR), and innovative approaches such as Multiple-Round *In Vivo* Site-Specific Assembly (MISSA) and Phage Enzyme-Assisted Direct Assembly (PEDA), are critically evaluated. The review highlights that method selection should consider the scale, complexity, cost, and specific needs of individual research projects, noting that no single technique is universally optimal. Future trends suggest the increased integration of enzymatic efficiency, host versatility, and automation, broadening the accessibility and capabilities of DNA assembly technologies.

**Word count:** 4,896 words, excluding references.

**Introduction**

DNA cloning encompasses various molecular techniques designed to precisely manipulate and engineer DNA molecules, enabling researchers to effectively assemble, edit, and restructure genetic sequences. Central to all these methodologies is the use of overlapping DNA sequences that allow for the precise joining of these molecules in a specific order. Some of these techniques employ the *in vivo* mechanism of homologous recombination (HR) to promote ligation of overlapping DNA sequences. In contrast, others have utilized an *in vitro* combination of different enzymes to generate large DNA constructs. Moreover, overlapping DNA sequences can be obtained using a combination of specific restriction DNA endonucleases, which are employed in conventional DNA cloning techniques or advanced methods, such as Biobricks or Golden Gate [1]. However, the effective use of restriction enzymes to generate overlapping sequences relies on the efficiency of the enzyme itself and the availability of the enzyme recognition site in the desired sequence [2]. To overcome these challenges, in 1990, Aslanidis and de Jong [3] proposed the first known ligation and restriction enzyme-independent cloning of DNA fragments by employing a combination of overlapping DNA sequences generated by polymerase chain reaction (PCR), followed by the treatment of those sequences with T4 DNA polymerase. From this pioneering work, different in vitro and in vivo restriction-free overlapping sequence cloning (RFOSC) techniques have been developed (Table 1). Thus, the primary purpose of this review was to compile the major *in vitro* and *in vivo* RFOSC techniques employed in synthetic biology to generate DNA molecule constructs from different overlapping sequences.

***In vitro* DNA RFOSC techniques**

*Gibson Assembly*

Gibson Assembly (Table 1) has emerged as a powerful alternative to conventional DNA cloning methods, enabling the generation of large DNA constructs, including the complete assembly of artificial bacterial genomes [4]. This method employs three key enzymes: a T5 bacteriophage-derived 5 exonuclease, a high-fidelity DNA polymerase (e.g., Phusion® DNA polymerase), and a thermostable Taq DNA ligase [5]. Successful assembly requires DNA fragments to be joined and share approximately 20-40 base pairs (bp) overlapping sequences at their 5' and 3' termini [6]. Gibson assembly is performed under isothermal conditions at 50 °C and typically takes approximately one hour to complete, and it does not require specialized equipment such as a thermocycler, making it accessible for a wide range of laboratory settings. This method allows for the assembly of DNA constructs hundreds of kilobases in size and a dozen different DNA parts [5,7]. Mechanistically, the process begins with the 5' exonuclease degrading the 5' ends of the DNA fragments, creating single-stranded complementary overhangs that facilitate annealing of homologous sequences. Subsequently, the high-fidelity DNA polymerase extends the 3' ends to fill any gaps, and the DNA ligase subsequently seals the nicks, resulting in a contiguous, fully assembled DNA molecule [8].

Gibson assembly achieves approximately 90% efficiency when generating plasmids of up to 15 kb in size [5]. However, the efficiency tends to decrease to approximately 50% when assembling more than four fragments or constructing plasmids larger than 4.8 kb [9]. High GC-content sequences also negatively impact assembly efficiency [10]. Moreover, the use of nonproofreading DNA polymerases during the assembly process has been associated with an increased frequency of single-nucleotide polymorphisms in the resulting DNA constructs [9].

*Circular Polymerase Extension Cloning (CPEC)*

The CPEC technique (Table 1) is a polymerase chain reaction (PCR)-based method that utilizes a DNA molecule, such as a gene, as a megaprimer to amplify a vector, thereby allowing direct cloning of the insert within the vector. This technique can be employed in gene libraries, high-throughput expression cloning, and multiway assembly of genetic pathways [11]. This cloning method utilizes PCR-generated overlaps between DNA fragments to facilitate seamless assembly of these fragments into circular plasmids. Specifically, CPEC relies on complementary regions at the ends of the insert and linearized vector, which are generated by carefully designed oligonucleotides during PCR amplification. These homologous regions serve as DNA polymerase templates, extending annealed fragments and synthesizing complementary strands to form double-stranded circular DNA molecules [11]. In addition, CPEC minimizes the risk of mutation accumulation owing to its limited amplification cycles, rendering it a highly accurate cloning strategy [12]. In practice, amplified DNA fragments are combined with linearized vectors, DNA polymerase, reaction buffer, or deoxyribonucleotide triphosphates (dNTPs). A single cycle of denaturation, annealing, and extension is sufficient for typical single-gene cloning applications. In contrast, assembly of gene libraries may require two–five cycles, whereas multi-gene constructs could necessitate five to twenty-five cycles to achieve optimal efficiency, potentially nearing 100% [11]. While this technique is suitable for generating medium-sized vectors of up to approximately 8 kb, intrinsic limitations of DNA polymerase and overlap-extension procedures typically restrict the efficiency of assembling vectors from numerous fragments or generating significantly larger constructs [13].

*Polymerase Incomplete Primer Extension (PIPE)*

PIPE cloning (Table 1) relies on deliberately generating partially single-stranded PCR products by omitting a complete extension step during amplification, resulting in variable 3' termini among the amplicons [14]. These single-stranded regions serve as overhangs that anneal through complementary 14-17 bp sequences introduced at the 5' ends of inserts and vectors via specific oligonucleotide design [15]. PIPE is a simple and efficient method that achieves up to 87% ligation efficiency without requiring extensive DNA purification beyond *Dpn*I treatment to eliminate template plasmid DNA [15]. In practice, the insert and vector were separately amplified by PCR to introduce homologous overlaps, then mixed and incubated for approximately 1 h at room temperature or overnight at 16°C, allowing for spontaneous annealing. Subsequent transformation into competent cells enables endogenous repair of the remaining nicks [14]. Notably, the PCR program for PIPE omits the final extension step, thereby increasing the frequency of partially extended molecules necessary for efficient cloning [15]. However, a limitation of PIPE is the generally low colony yield post-transformation, which can restrict its application in library construction and the possibility of background colonies resulting from the re-circularization of the vector backbone without an insert [16].

*Overlap Extension Cloning (OEC)*

Overlap Extension Cloning (OEC; Table 1) utilizes a DNA insert as a mega-oligonucleotide to mediate its incorporation into a target vector. The 5' and 3' ends of the insert were PCR-amplified with sequences homologous to the vector flanking regions, which correspond precisely to the insertion site of the insert after ligation [17]. During the second PCR reaction, both the vector and the insert were denatured, and the single-stranded insert served as a large primer, annealing the vector at the homologous regions, and priming DNA synthesis in both directions. This process generates a circular double-stranded DNA molecule with a single nick on each strand [17]. OEC is a rapid method, achieving efficiencies of up to 90% when inserting fragments smaller than 1 kb; however, efficiency declines to approximately 45% when fragments larger than 4 kb are used [16]. To maximize efficiency, designing extensive homology arms of at least 50 base pairs is critical, ensuring a high GC content, and using primers with a melting temperature (Tm) above 65 °C [18]. Additionally, it is recommended to use an annealing temperature of 5-10 °C below the standard calculated Tm. While a 1:5 vector-to-insert molar ratio is typically sufficient, using a ratio of 1:250 can further improve the cloning efficiency [17]. Phusion® DNA polymerase is preferred for the second PCR step because its lack of strand displacement activity is essential for accurately generating recombinant molecules [18]. However, OEC are highly sensitive to the presence of primer dimers, which can inhibit efficient amplification. Therefore, careful primer design is necessary. Alternatively, a pretreatment step with T4 DNA polymerase in the absence of dNTPs can be employed to selectively degrade primer-dimers via the enzyme's 3'→5' exonuclease activity, thereby improving the quality of insert preparation and subsequent amplification [16].

*Flap Endonuclease Cloning (FEN-Cloning)*

FEN-Cloning (Table 1) encompasses a family of methods that exploit the activity of flap endonucleases, particularly uracil DNA glycosylase (UDG), to facilitate the efficient and scarless assembly of DNA fragments [19,20]. By enabling the simultaneous insertion of up to seven DNA sequences while maintaining seamless junctions between parts [21], FEN-Cloning has applications in molecular cloning, nucleic acid and protein engineering, and the introduction of site-specific mutations [22]. In FEN-based assembly, deoxyuridine residues are strategically incorporated into PCR-amplified fragments and vectors at the flanking regions of 6-12 bp homologous overlaps. These residues were introduced near the 3' end of the primers used during amplification. After PCR, treatment with a combination of UDG and an exonuclease (typically Exonuclease VIII or Endonuclease IV) cleaves the uracil sites, generating single-stranded overhangs that expose the homologous sequences [23]. Assembly proceeded by simple incubation at 37 °C for approximately 15 min, promoting the annealing of complementary overhangs between fragments. The optimal length of homologous sequences depends on the A/T content, with 8-10 bp overlaps recommended for sequences exhibiting approximately 50% A/T content [23]. A standard molar ratio of 1:3 (vector to each insert) was recommended for the reaction setup, with all fragments present at equimolar concentrations. In addition to being dependent on the incorporation of uracil residues, FEN-Cloning is enzyme-specific; DNA polymerases capable of reading through uracil, such as *Taq* and *Pfu* polymerases, must be used during PCR amplification to avoid stalling [24]. Following initial assembly at 37 °C, a secondary incubation at room temperature for 15 min can enhance complete ligation of residual overhangs [21], leading to overall cloning efficiencies between 90% and 98%. Moreover, improvements to FEN-based methods include engineered endonuclease enzymes that combine uracil recognition and nicking activity into a single thermostable protein, thus streamlining the process and reducing the number of enzymatic steps. These enhanced versions improve the robustness of the technique, allowing high-efficiency assembly of larger constructs, minimizing unwanted exonucleolytic degradation, and expanding the utility of the method for complex synthetic biology applications, including modular genome construction and metabolic pathway engineering [25].

**Hybrid RFOSC techniques**

*Ligation-Independent Cloning (LIC)*

LIC (Table 1) leverages the exonuclease and polymerase activities of the T4 DNA polymerase enzyme to enable seamless assembly of DNA fragments without the need for traditional ligation [3,26]. In LIC, vectors are pre-designed *in silico* to contain a specific region where a given nucleotide is intentionally absent, precisely at the intended insertion site [3]. The vector backbone is treated with T4 DNA polymerase in the presence of only the dNTP corresponding to the missing nucleotide, whereas the DNA insert, flanked by appropriate adapter sequences, is treated similarly to complementary dNTP [27]. Because of the 3′→5′ exonuclease activity of T4 DNA polymerase in the absence of dNTPs, single-stranded complementary overhangs are generated on both the vector and insert, facilitating ligation-independent cloning methods [3]. Upon encountering the first nucleotide matching the supplied dNTP, the enzyme switches to polymerase activity, thus limiting overdigestion and preserving the designed overlaps. This facilitates precise annealing between the insert and vector, leading to circularization [28]. Vectors specifically designed for LIC are available [27,29], which simplifies the workflow for synthesizing the desired DNA fragment. LIC offers high cloning accuracy, with reported efficiencies ranging between 75 and 90%, and the assembly reaction itself can be completed in approximately 30 min [27], excluding preparation and transformation times. This method is particularly advantageous for projects requiring insertion of multiple amplification products into a small number of vectors [30]. However, a limitation is the dependency on predesigned vectors or the need to synthesize custom vectors de novo, which constrains the range of sequences that can be conveniently cloned [31].

*Sequence and Ligation-Independent Cloning (SLIC)*

The SLIC technique (Table 1) eliminates the need for predesigned vectors by relying on sequence-independent assembly to optimize traditional ligation-independent cloning methods. Both the DNA fragments and vectors were amplified by PCR using primers that created specific homologous sequences at their ends. After amplification, the fragments were separately treated with T4 DNA polymerase and appropriate nucleotides at room temperature for 30-60 min, allowing the exonuclease activity of the enzyme to generate single-stranded homologous regions. Subsequently, the treated fragments and vectors were combined and incubated at 37 °C for 30 min to promote overlap and annealing before transformation [32]. More recent optimizations, such as those by Islam et al. [33], demonstrate that SLIC can be performed in a single step by incubating the DNA mixture for 30 s at 50 °C using homology regions shorter than 20 base pairs. In this optimized protocol, 0.3 U of T4 DNA polymerase is used per 100 ng of vector DNA with a recommended vector-to-insert molar ratio of 1:2. Under these conditions, up to four fragments can be concatenated efficiently into a vector per reaction. However, efficiency tends to decrease as fragments increase [34].

*T5 Exonuclease-Dependent Assembly (TEDA)*

TEDA (Table 1) is a simplified Gibson assembly-like method that uses a single enzyme. Xia et al. [35] demonstrated that T5 exonuclease alone, when added to a solution containing Tris-HCl, $MgCl_2$, and 8000, was sufficient to mediate DNA fragment assembly and generate a circular vector. TEDA is a one-step, one-pot method that achieves assembly efficiencies greater than 95%, with a total reaction time of approximately 40 min and a cost of less than one dollar per reaction. In TEDA, vectors are linearized by restriction digestion or by PCR amplification, whereas inserts are PCR-amplified to introduce homologous regions at their ends. The components were mixed with the TEDA solution in vector-to-insert ratios of 1:1 to 1:4. After a brief incubation, the mixture was directly transformed into competent bacteria, where endogenous repair mechanisms complete the assembly. The TEDA method has proven efficient and cost-effective for assembling small-to moderate-sized DNA constructs [35]. However, systematic studies evaluating its scalability for multifragment or large construct assemblies are still limited, and its performance in such contexts requires further empirical validation.

*T5 exonuclease–mediated low-temperature sequence- and ligation-independent cloning method (TLTC)*

TLTC is a recently introduced method designed for low-cost scarless DNA assembly using a single enzyme under simplified conditions. Developed by Yu et al. [36], TLTC harnesses the residual exonuclease activity of T5 at 0 °C to gently digest DNA ends, producing short single-stranded overhangs (~15-25 nucleotides). This allows for efficient annealing between overlapping sequences of the vector and insert(s), followed by transformation into *E. coli,* where gap repair is completed *in vivo.* TLTC eliminates the need for DNA ligase or polymerase and operates at low temperatures without the need for thermal cycling or proprietary reagents. It achieves >95% efficiency in single-fragment cloning and ~80% efficiency for four-part assemblies, with comparable or superior performance to Gibson Assembly under short-overlap conditions. Additionally, TLTC can tolerate and eliminate non-homologous vector termini, making it a versatile and accessible tool for routine cloning tasks in resource-limited settings [36].

*In-Fusion™ Assembly*

In-Fusion™ Assembly (Takara Bio) (Table 1) is a commercially available kit-based method for seamless DNA cloning. The main advantage of In-Fusion™ is the convenience of ready-to-use reagents, although this technique is relatively expensive. Despite the standardized commercial format, several variables can be adjusted, including primer length and overlapping region size, achieving efficiencies between 60% and 85% [37]. Typically, oligonucleotides are designed to amplify fragments with at least 15 bp homology to the target vector [38]. Longer overlapping regions tend to increase the assembly efficiency. Although the oligonucleotide size did not significantly affect the final efficiency, the number of inserted fragments was inversely correlated with assembly success. In addition, this technique uses a proprietary enzyme blend that includes a 5′→3′ exonuclease activity functionally similar to T5 exonuclease to generate short single-stranded overhangs at the ends of the DNA fragments. These overhangs enable specific annealing of homologous sequences between the vector and insert(s). The remaining gaps and nicks are repaired after transformation by the host's endogenous recombination and DNA repair machinery [37]. In-Fusion™ cloning is well suited for complex applications, including assembling two or more BioBricks in a single reaction and mutagenesis workflows. While it enables the assembly of multiple fragments simultaneously, it does not inherently support hierarchical cloning strategies where multiple assembly steps are integrated into a single iterative process [39].

*Flp Double Cross System*

Although initially developed for genome editing, the Flp Double Cross System (Table 1) can be employed in plasmid DNA assembly, particularly in homology-based cloning strategies. It utilizes the Flp recombinase, an enzyme from *Saccharomyces cerevisiae*, which recognizes specific FRT (Flp Recognition Target) sites composed of short palindromic sequences separated by an asymmetric core [40]. The FLP/FRT system enables site-specific recombination by recognizing FRT sites in both the vector and insert, catalyzing a double-reciprocal crossover event that facilitates the exchange of genetic material at defined genomic loci [41]. A significant advantage of the Flp system is that it does not require any restriction enzymes or ligases. It offers a highly versatile alternative for cloning DNA fragments that are difficult to handle using conventional methods, including PCR products with problematic 3'-termini. Additionally, Flp-mediated recombination allows the generation of clones with fragments in either orientation, enhancing flexibility in plasmid construction, and enabling multiple fragment assemblies [42].

*TOPO® Cloning*

TOPO® cloning (Table 1) is a rapid single-enzyme technique for inserting DNA fragments into plasmids without ligases. It employs the properties of topoisomerase I, which acts as an endonuclease and ligase to insert PCR-amplified fragments into linearized vectors [43]. The vectors were pre-activated with covalently bound topoisomerase at specific (T/C)CCTT recognition sequences at their ends [43], facilitating direct cleavage and ligation of DNA inserts possessing either single-base overhangs (produced by *Taq* DNA polymerase) or blunt ends (generated by proofreading polymerases). This single-step reaction was completed within approximately five minutes at room temperature and achieved cloning efficiencies exceeding 85%. Although relatively expensive, TOPO cloning is highly valuable for quick multi-gene cloning, high-throughput applications, and downstream workflows, including large-scale cloning, vector construction, and cDNA expression studies in mammalian systems [44].

## *In Vivo* RFOSC

### *In Vivo Assembly (IVA)*

*In vivo* assembly (IVA; Table 1) takes advantage of the RecA-independent recombination pathways present in laboratory strains of *Escherichia coli* [45]. Although the precise molecular mechanisms remain incompletely understood, these pathways enable efficient recombination of linear DNA fragments, likely through an annealing-based process. Unlike RecA-mediated recombination, which requires 150-300 bp of homology, RecA-independent recombination functions with shorter homology arms (~20-40 bp), allowing for more straightforward cloning procedures [46]. IVA allows for insertions, deletions, mutagenesis, and subcloning, using a universal protocol. The workflow involves single-tube PCR amplification of DNA sequences, followed by *Dpn*I digestion (1-2 h) to remove the template DNA and transform it into competent *E. coli*. Homology-based recombination in vivo assembles the final construct with efficiencies approaching 90% under optimized conditions [45]. However, it has been reported that only up to five DNA fragments can typically be assembled in a single reaction, and the efficiency heavily depends on the competency of the bacterial host strain [47]. Finally, simplified IVA-like protocols have been developed, including the AQUA cloning technique [48], which allows the construction of multipart DNA molecules.

*Multiple-Round In Vivo Site-Specific Assembly (MISSA)*

MISSA (Table 1) is an efficient in vivo DNA assembly technique designed primarily for multi-gene transformation of plants. It relies on the mating-assisted genetically integrated cloning (MAGIC) technique employed for the generation of recombinant DNA [49] with the advantage of allowing the assembly of multiple transgenes or even genetic pathways [50]. MISSA employs bacterial conjugative transfer and two site-specific recombination events mediated by Cre recombinase and λ phage recombination proteins, achieving efficiencies of up to 60% [50]. Cre recombinase recognizes and catalyzes recombination between 34 bp loxP sequences, whereas λ integrase (Int) and excisionase (Xis) catalyze recombination between bacterial and phage att sites, generating junctions between host and donor DNA. Each round of MISSA involves the initial integration of the donor plasmid into the recipient plasmid via Cre/loxP recombination, followed by removal of the donor backbone via λ phage-mediated recombination. Up to 15 iterative rounds can be used to assemble complex, multi-gene constructs. These limitations include the requirement for specially designed vectors and potential expression issues caused by att-site insertions [50].

*Phage Enzyme-Assisted Direct In Vivo DNA Assembly (PEDA)*

PEDA (Table 1) is an *in vivo* DNA assembly method that expresses phage-derived enzymes inside the host cell to assemble the DNA fragments. In PEDA, T5 exonuclease and T4 DNA ligase coding sequences are co-delivered into the host with the DNA parts of interest [51]. Once expressed, these enzymes mediate DNA end-processing and ligation *in vivo,* enabling the assembly of DNA constructs using homologous regions as short as five bp. PEDA has been successfully applied across diverse non-conventional hosts, including *Pseudomonas*, *Lactobacillus*, *Ralstonia*, and *Yarrowia lipolytica*, demonstrating its broad applicability. Overexpression of endogenous RecA and Gam recombinases in *E. coli* further enhanced the PEDA efficiency. This method represents a cost-effective and straightforward alternative for complex DNA assemblies in various organisms.

*Yeast Homologous Recombination (YHR)*

YHR (Table 1) uses the highly efficient endogenous homologous recombination machinery of *Saccharomyces cerevisiae* for *in vivo* DNA assembly. It allows the assembly of up to 12 DNA fragments with as little as 24 bp of homologous overlap in a single reaction [52,53]. DNA assembly using YHR involves the co-transformation of linearized vectors and/or PCR fragments into yeast cells, where the yeast HR machinery generates three single-stranded overhangs that invade homologous DNA regions, forming Holliday junctions, gap filling, and ligation [52,53].

YHR has been successfully used in synthetic chromosome construction, genome refactoring, and metabolic pathway engineering and is particularly powerful for assembling large constructs (>100 kb) and repetitive sequences that are problematic in bacterial systems [4].

*Transformation-Associated Recombination (TAR)*

The TAR (Table 1) and YHR share the same core homologous recombination machinery. However, TAR is tailored to isolate and assemble large genomic regions (up to 300 kb) without the need for genomic libraries [54]. In TAR, a linearized vector bearing two "hooks" (homologous sequences ~60 bp) captures the target DNA directly from the genome during yeast transformation. TAR achieves cloning efficiencies nearly 100,000-fold higher than traditional methods, such as YAC or BAC, with positive clone frequencies of approximately 4% to 20%, employing suicidal vectors expressing the killer toxin K1 [55]. However, TAR and YHR require yeast-specific shuttle vectors and are less flexible than bacteria-based cloning platforms, posing practical limitations in many standard molecular biology laboratories.

**Discussion**

The rapid development of recombination-based fragment-overlap sequence cloning (RFOSC) methods has significantly expanded the molecular biology toolbox, enabling more versatile, efficient, and cost-effective DNA manipulation strategies [5,32]. Although these techniques share a common foundation, that is, the exploitation of restriction-free overlapping sequences, they differ markedly in complexity, efficiency, scalability, and applicability. Gibson Assembly remains the gold standard, renowned for its high efficiency and robustness, particularly in large-scale assemblies and synthetic genome construction [4]. However, its relatively high cost and reliance on commercial reagents may limit its routine use in low-budget laboratories, particularly when multiple large assemblies are required. In this context, techniques such as TEDA [35], TLTC [36], and SLIC [49] provide excellent alternatives, offering comparable efficiencies while reducing the cost and procedural complexity. In contrast, TOPO Cloning [43] and In-Fusion Assembly [37] deliver high accuracy and user-friendliness. However, they depend heavily on proprietary enzyme systems, raising accessibility concerns, particularly in laboratories with limited financial resources. Meanwhile, simpler kit-free approaches such as PIPE [14], CPEC [13], and OEC [18] remain attractive options, offering effective cloning without requiring specialized reagents or complex workflows. A major distinction emerged when comparing *in vitro* enzymatic methods with *in vivo* recombination-based systems. In vivo techniques, such as *In Vivo* Assembly (IVA) [45] and Yeast Homologous Recombination (YHR) [4], exploit the natural recombination pathways of the host organism to assemble DNA fragments, reducing reliance on in vitro enzymatic reactions and specialized equipment. These approaches are particularly well-suited for modular cloning frameworks in synthetic biology and metabolic engineering. However, their efficiency can be limited in organisms where the homologous recombination machinery is less efficient or poorly characterized. Notably, emerging systems, such as Phage Enzyme-Assisted Direct Assembly (PEDA) [51], are being adapted beyond their original genomic integration purposes. These innovations are expanding the possibilities for direct DNA assembly across various non-model organisms, further pushing the boundaries of traditional recombination-based methodologies.

The selection of a DNA assembly strategy should be carefully aligned with the target application, the desired scale, available infrastructure, and budgetary constraints. A clear understanding of each method's strengths and limitations is crucial for its successful implementation in diverse research contexts, ensuring that the chosen system meets both the technical demands and practical realities of the intended project.

**Conclusion**

The growing diversity of DNA assembly methods has revolutionized molecular biology, synthetic biology, and genetic engineering. Techniques such as Gibson Assembly, TEDA, SLIC, PIPE, CPEC, OEC, In-Fusion™, FEN-Cloning, TOPO® Cloning, and emerging systems such as PEDA, MISSA, and *in vivo* recombination strategies (*e.g.,* IVA, YHR, and TAR) have collectively expanded the capacity for efficient, modular, and cost-effective DNA manipulation. Each method offers unique advantages (Table 1), such as speed and simplicity *(e.g.,* TOPO®, PIPE, TEDA). Others emphasize accuracy and seamless scarless assembly (e.g., Gibson Assembly, SLIC, In-Fusion™), while in vivo systems (*e.g.,* IVA, YHR, TAR) harness cellular machinery for the assembly of large and complex constructs. Despite the variety of available techniques, no "universal" method is suitable for all applications. The optimal choice depends on the size and number of DNA fragments, the target organism, intended downstream application, cost constraints, and available laboratory infrastructure. Understanding the strengths and limitations of each approach is critical for tailoring strategies to achieve specific experimental goals. Future developments in DNA assembly are expected to integrate the best features of both *in vitro* and *in vivo* methodologies by combining efficiency, scalability, cost-effectiveness, and host versatility. Advances in enzyme engineering, automation, and synthetic biology frameworks will likely further democratize the access to high-performance DNA assemblies. In this evolving landscape, strategic selection and combination of methods are key to accelerating research across diverse fields, from fundamental biology to large-scale genome engineering.

**Funding**

This work was supported by the Conselho Nacional de Desenvolvimento Científico e Tecnológico – CNPq [grant number 314558/2020-9], by "Programa Pesquisador Gaúcho - PqG" from Fundação de Amparo à Pesquisa do Estado do Rio Grande do Sul - FAPERGS [Edital FAPERGS 07/2021 and grant number 21/2551-0001958-1], and by "Programa de Redes Inovadoras de Tecnologias Estratégicas do Rio Grande Do Sul – RITEs-RS" from Fundação de Amparo à Pesquisa do Estado do Rio Grande do Sul - FAPERGS [Edital FAPERGS 06/2021 and grant number 22/2551-0000397-4]. The sponsors had no role in the study design, collection, analysis, interpretation of data, writing of the report, or the decision to submit the article for publication.

**Declaration of generative AI and AI-assisted technologies in the writing process**

During the preparation of this study, we used ChatGPT (https://chatgpt.com) and PaperPal (https://paperpal.com) to improve the readability and language of the manuscript. After using these tools and services, the authors reviewed and edited the content as needed, and took full responsibility for the content of the published article.

**Author contributions**

**Isabella Frighetto Bomfiglio:** Conceptualization, Writing - Original Draft, Writing - Review & Editing, Visualization. **Isabelli Seiler de Medeiros Mendes:** Conceptualization, Writing - Original Draft, Writing - Review & Editing, Visualization. **Diego Bonatto:** Conceptualization, Writing - Original Draft, Writing - Review & Editing, Visualization, Supervision, Project Administration, Funding Acquisition.

**Declaration of interest**

Declarations of interest: none.

**Tables**

Table 1. Major restriction-free overlapping sequence cloning (RFOSC) employed for DNA assembly.

| Technique | Type | Number of fragments | Main Reagents | Reaction Time | Efficiency (%) |
|---|---|---|---|---|---|
| Gibson Assembly | *In vitro* | 5–6+ | Exonuclease, Pol, Ligase | ~1 h | >95% |
| CPEC | *In vitro* | 3–5 | DNA Polymerase | 1–25 PCR cycles | Up to 100% |
| PIPE | *In vitro* | 1 or mutations | PCR, truncated primers | 1 h – overnight | ~87% |
| FEN-Cloning | *In vitro* | Up to 7 | Uracil-DNA Glycosylase, dU | ~15–30 min | 90–98% |
| OEC | *In vitro* | 1 | PCR, DNA as primer | ~2–3 h | High |
| SLIC | *Hybrid* | 4–5 | T4 DNA Pol, PCR | ~30 min – 1 h | High |
| LIC | *Hybrid* | 1 (multi rare) | T4 DNA Pol, selected dNTPs | ~30 min | 75–90% |
| TEDA | *Hybrid* | 1–2 | T5 Exonuclease, buffer | ~40 min | >95% |
| TLTC | *Hybrid* | 1-4 | T5 exonuclease | 5 min | >95% (1 fragment), ~80% (4 fragments) |
| In-Fusion | *In vitro* | 1–4 | Commercial enzyme mix | ~15–30 min | 60–85% |

| Technique | Type | Number of fragments | Main Reagents | Reaction Time | Efficiency (%) |
|---|---|---|---|---|---|
| TOPO Cloning | *In vitro* | 1 | Topoisomerase-linked vector | ~5 min | >85% |
| IVA | *In vivo* | Multiple | PCR, E. coli | PCR + transformation | High |
| MISSA | *In vivo* | Multiple | Conjugation, recombinases | Overnight | High |
| PEDA | *In vivo* | Multiple | Phage integrases | Variable | High |
| YHR | *In vivo* | Multiple | Yeast recombination machinery | Variable | High |
| TAR Cloning | *In vivo* | Large genomic regions | Yeast with homology arms | Variable | High |
| Flp System | *In vivo* | 1 | Flp recombinase, FRT sites | Variable | High |